\documentclass[12pt]{iopart}

\usepackage{iopams}
\usepackage{graphicx}
\usepackage[colorlinks,
                   linkcolor=blue,
                  anchorcolor=red,
                  citecolor=green
                  ]{hyperref}

\newcommand{\beqn}{\begin{eqnarray}}
\newcommand{\eeqn}{\end{eqnarray}}
\newcommand{\lb}{\label}
\newcommand{\beq}{\begin{equation} }
\newcommand{\eeq}{\end{equation} }
\newcommand{\fr}{\frac}
\newcommand{\al}{\alpha}
\newcommand{\be}{\beta}
\newcommand{\ka}{\kappa}
\newcommand{\om}{\omega}
\newcommand{\ga}{\gamma}
\newcommand{\de}{\delta}
\newcommand{\ep}{\epsilon}

\newcommand{\Ga}{\Gamma}
\newcommand{\po}{\mbox{\boldmath $\omega$}}
\newcommand{\ptau}{\mbox{\boldmath $\tau$}}
\newcommand{\ppsi}{\mbox{\boldmath $\psi$}}
\newcommand{\ps}{\mbox{\boldmath $\sigma$}}
\newcommand{\pxi}{\mbox{\boldmath $\xi$}}
\newcommand{\pGamma}{\mathbf \Gamma}
\newcommand{\bG}{\mathbf  G}
\newcommand{\bI}{\mathbf  I}
\newcommand{\bT}{\mathbf  T}
\newcommand{\pF}{\textbf{\emph{F}}}
\newcommand{\pI}{\textbf{\emph{I}}}
\newcommand{\pS}{\textbf{\emph{S}}}
\newcommand{\pU}{\textbf{\emph{U}}}
\newcommand{\px}{\textbf{\emph{x}}}
\newcommand{\pu}{\textbf{\emph{u}}}
\newcommand{\pf}{\textbf{\emph{f}}}
\newcommand{\pn}{\textbf{\emph{n}}}
\newcommand{\pe}{\textbf{\emph{e}}}

\newcommand{\pa}{\textbf{\emph{a}}}
\newcommand{\pgg}{\textbf{\emph{g}}}
\newcommand{\pat}{\partial}
\newcommand{\na}{\nabla}
\def\Xint#1{\mathchoice
{\XXint\displaystyle\textstyle{#1}}%
{\XXint\textstyle\scriptstyle{#1}}%
{\XXint\scriptstyle\scriptscriptstyle{#1}}%
{\XXint\scriptscriptstyle\scriptscriptstyle{#1}}%
\!\int}
\def\XXint#1#2#3{{\setbox0=\hbox{$#1{#2#3}{\int}$}
\vcenter{\hbox{$#2#3$}}\kern-.5\wd0}}

\def\dashint{\Xint-}

\begin{document}

\title[Zonal structure of unbounded external-flow and aerodynamics]{Zonal structure of unbounded external-flow and aerodynamics}

\author{L. Q. Liu, L. L. Kang and J. Z. Wu}

\address{State Key Laboratory of Turbulence and Complex System, Center for Applied Physics and Technology, College of Engineering,
Peking University, Beijing 100871, China}

\ead{lqliu@pku.edu.cn}

\begin{abstract}
This paper starts from the far-field behaviours of velocity field in externally-unbounded flow. We find that the well-known algebraic decay of disturbance velocity as derived kinematically is too conservative. Once the kinetics is taken into account by working on the fundamental solutions of far-field linearized Navier-Stokes equations, it is proven that the furthest far-field zone adjacent to the uniform fluid at infinity must be unsteady, viscous and compressible, where all disturbances degenerate to sound waves that decay exponentially. But this optimal rate does not exist in some commonly used simplified flow models, such as steady flow, incompressible flow and inviscid flow, because they actually work in true subspaces of the unbounded free space, which are surrounded by further far fields of different nature. This finding naturally leads to a zonal structure of externally-unbounded flow field. The significance of the zonal structure is demonstrated by its close relevance to existing theories of aerodynamic force and moment in external flows, including the removal of the difficulties or paradoxes inherent in the simplified models.
\end{abstract}

\vspace{2pc}
\noindent{\it Keywords}: External flow; Far-field velocity; Zonal structure; Aerodynamics

\maketitle

\section{Introduction}\label{sec.Introduction}

A fundamental issue in all studies of externally unbounded flows is the asymptotic
behaviour of velocity field as $r \equiv |\px| \rightarrow \infty$ ($\px$ is the position vector).
This is a necessary prerequisite for not only prescribing far-field boundary conditions for external-flow problems, but also ensuring the convergence of relevant integrals over the entire externally unbounded space with the fluid in uniform state at infinity (below we use the word `free space' for short) or arbitrarily large external boundary. This issue has long been an important subject for mathematicians working on the existence and uniqueness of the solutions of Navier-Stokes (NS) or Euler equations. But in the field of applied fluid dynamics and aerodynamics, when conducting the theoretical analysis or numerical computation of a specific external-flow problem, one seldom asks, in that problem, what the `infinity' means (does it really reach the uniform fluid at the `true' infinity?) and what the decay rates of various disturbances are (do they decay algebraically as $O(r^{-m})$ or exponentially as $O(e^{-r^k})$, with $m,k>0$?). The answer to these questions actually varies from one flow model to another, for example from steady to unsteady flows, from incompressible to compressible flows, and from inviscid to viscous flows. In the present paper we address this fundamental issue by using far-field analysis, both kinematically and kinetically. Here, kinematics is referred to as the study of motions of themselves apart from considerations of mass and force, and kinetics is referred to as the study of changes of motions produced by forces (Webster 1953). The results provide a precise estimate of the disturbance decay rates and associated concept of `infinity' as used in different flow models.

Specifically, it is well known that if the vorticity and dilatation fields are physically compact (i.e., they are significant only in a finite zone, say $V_{\rm c}$, outside which they decay exponentially), then the far-field velocity must be irrotational and incompressible, which decays only algebraically (Batchelor 1967). But the assumed decay rate of vorticity and dilatation itself needs a proof. This can only be done by entering kinetics. As one of the major findings of the present paper, we shall not only confirm that assumption, but also further prove that the algebraic decay rule of velocity only represents an upper bound and is too conservative. Kinetically, it can be sharpened to exponential decay of all disturbance quantities (including velocity), if and only if the flow is {\it unsteady, viscous and compressible}. Evidently, exponential decay ensures a smooth transition to uniform fluid status at infinity and the convergence of various integrals over arbitrarily large domain.

Unfortunately, all other simplified flow models (e.g., incompressible flow, steady flow, and inviscid flow, etc.) do not enjoy the nice feature of exponential decay. But this situation by no means implies that unsteady, viscous and compressible flow is an exception; rather, it just reflects the inherent physical incompleteness of those models, namely they only work in some true subspaces of the free space. Thus, in these models when one talks about some far-field conditions as $r \rightarrow \infty$, the `infinity' there cannot really reach the uniform fluid at infinity in free space, while only the `infinity' in unsteady, viscous and compressible flow can. This observation naturally leads to a physical picture of the far fields for the first time: they form a zonal structure. Some long-standing or new puzzles or paradoxes concerning far-field behaviour can thereby be well clarified.

In \Sref{sec.Dynamics} we develop the far-field analysis and derive the desired decay rates, starting from the linearized NS equations for general unsteady, viscous and compressible flow. The zonal structure of far fields for different flow models are presented in \Sref{sec.zonal}. The close relevance of this structure to theories of aerodynamic force and moment on a body moving in externally unbounded flow is discussed in \Sref{sec.force}, followed by conclusions of \Sref{sec.Conclusion}. To be self-contained, a kind of inverse Laplace transform is reproduced from Lagerstrom \etal (1949) in \ref{appendix:Laplace}.

\section{Far-field asymptotics and decay rate}\label{sec.Dynamics}

The decay behaviour of far-field velocity in externally unbounded domain, denoted by $V_\infty$ here and after, can be studied kinematically and kinetically. To orient our approach, it is necessary to compare these two methods first. This is done for the first time in \S~\ref{subsec.kinematics}, which shows that only kinetic method can obtain a complete description of the far-field decay rate. Then, the linearized NS equations are reduced in \S~\ref{subsec.NS}, of which the fundamental solutions of decoupled type and coupled type are carefully studied in \S~\ref{subsec.decoupled} and \S~\ref{subsec.coupled}, respectively, both proving that the far-field disturbances decay exponentially and the result obtained by kinematic method is only an upper bound. This result is crucial for the constructing of a complete zonal structure.

\subsection{From kinematics to kinetics}\label{subsec.kinematics}

Batchelor (1967) has given a clear presentation of the kinematic method for estimating the far-field velocity decay rate. By the classic Helmholtz decomposition of velocity field
\begin{equation}\lb{u-lt-1}
  \pu = \pu_\phi+\pu_\psi = \na\phi +\na\times \ppsi, \quad
 \na \cdot \ppsi =0,
\end{equation}
where $\phi$ and $\ppsi$ are scalar and vector potentials, respectively, one has Poisson's equations
\begin{equation}
  \vartheta \equiv \na\cdot \pu = \na^2\phi, \quad
  \po \equiv \na\times \pu = -\na^2\ppsi.
\end{equation}
Their fundamental solutions in $V_\infty$ yield the familiar generalized Biot-Savart law (Wu 2005, Wu \etal 2006, Wu \etal 2015), to which Batchelor applied the Taylor expansion and proved that the far-field flow must be incompressible and irrotational, which decays algebraically in space. The only assumption is that $\po$ and $\vartheta$ should be physically compact, so that the Taylor expansion is convergent.
This assumption itself was not proved. Nevertheless, since this method is purely kinematic, it has been regarded as universally true, no matter whether the flow is steady or unsteady, compressible or incompressible, laminar or turbulent, etc.

In our view, however, this kinematic estimate can hardly describe the true far-field asymptotics of externally unbounded flow. Rather, it only gives an upper bound of the velocity decay rate and is too conservative, because the Poisson equation only considers the spatial effect without any temporal evolution of the flow field where kinetic effects such as viscosity must enter. To obtain the optimal estimate of the far-field asymptotic decay rate, therefore, it is necessary to go into kinetics, based on the NS equations \eref{NSeqs-ch2} below.

The first effort toward this goal was made by Wu (1982), who used the vorticity transport equation of incompressible flow to prove that $\po$ is indeed compact as Batchelor (1967) assumed. He then used the Biot-Savart law to confirm Batchelor's estimate: $\pu$ decays as $r^{-n}$, where $n=2,3$ is the dimensionality of the space. Following the same strategy of Wu (1982), Liu \etal (2014) proved that $\po$ and $\vartheta$ of compressible flow must be compact.
Then, using the generalized Biot-Savart law they obtained the same result as that for incompressible flow. However, for compressible flow this result is still not optimal: for an observer standing outside the reach of the first sound wavefront generated by the body motion, the fluid should keep undisturbed and thus the velocity decays arbitrarily fast, say exponentially.

Instead of appealing to Poisson's equation, the most thorough approach to this problem would be directly deriving the far-field asymptotics from the NS equations. But, since the flow is critically dependent on the initial and boundary conditions, for example, under the same nonlinear NS equations with specified initial and boundary conditions the flow can be either laminar or turbulent, it is impossible to determine the far-field behaviour by the full NS equations. To bypass this difficulty, we assume that there exists a far-field zone neighbouring the uniform fluid at infinity, where the governing equations can be linearized and the effects of the initial and boundary conditions can be mimicked by proper source terms. Thus, the fundamental solution of the linearized NS equations obtained by Lagerstrom \etal (1949) can be applied. Intuitively, the assumed existence of linearized far field should be a physical fact since the flow at infinity always recovers to the uniform state and before that the disturbances of the flow must have decayed arbitrarily small. Actually, Lagerstrom (1964) has clearly stated that a linear zone should exist in \textit{viscous flow} around a \textit{finite-size} object. Although this assertion has not been mathematically proven for general NS flow, in our case the assumed existence of linear far field can be checked after the linearized solutions are obtained.

At this stage, the expressions of the equivalent source terms need to be carefully treated. This is relatively simple for steady flow, where the mass and energy sources are absent and the momentum source can be simplified to an impulsive force of unit mass fixed on the space,
\begin{equation}\label{f-st}
  \pf =-\fr{\de(\px)}{\rho_0}\pF,
\end{equation}
where $\pF$ is the total force experienced by the body, $\delta(\px)$ is the Dirac delta function, and suffix 0 denotes the uniform constant value at infinity. Under these assumptions, Liu \etal (2015) have obtained the far-field asymptotic expression of the velocity in two-dimensional (2D) steady flow region, say $V_{\rm st}$, which decays algebraically for both incompressible and compressible flows. In particular, both in the vortical wake and along the primary shock waves, the disturbance velocity $\pu'$ decays as $r^{-1/2}$. This estimate can by no means be reached by purely kinematic method, since the steady vortical wake must extend to downstream and finally escape from $V_{\rm st}$, making $\po$ noncompact. The same approach has been applied to three-dimensional (3D) steady flow, by which we found that $\pu'$ decays as $r^{-1}$ in the vortical wake and $r^{-5/4}$ along the primary shock waves (Liu 2016). However, for more general case where the object is allowed to move and deform arbitrarily, the expressions of the source terms can hardly be obtained.

This difficulty is likely associated with the very fact that in the formula for total force $\pF$ there must be a volume integral of $(\rho \pu)_t$ due to local flow unsteadiness (see \Eref{NSeqs-ch2b} below), making it impossible to express $\pF$ by boundary integrals only, which however is the prerequisite of expressing $\pF$ by linearized far-field variables. But incompressible flow is a pleasant exception, where for calculating the force (not the moment) one has transformation (Saffman 1992, Noca \etal 1999, Wu \etal 2005)
\begin{equation}
  \int_V\pu_t \mathrm{d}V = \int_{\pat V}\px (\pu_t\cdot \pn) \mathrm{d}S
  \quad {\rm if} \ \na\cdot \pu =0.
\end{equation}
Nevertheless, this limitation for flow incompressibility occurs only in the study of aerodynamic force. It does not affect the compactness of sources in estimating the far-field decay rate of unsteady compressible and viscous flow. For the latter purpose, the fundamental solution of linearized equations can still lead to the desired correct results.

\subsection{The linearized NS equations}\label{subsec.NS}

Consider a body $B$ moving and deforming arbitrarily in a canonically perfect gas externally unbounded and at rest at infinity. The continuity equation, NS equation per unit volume, and energy equation read (Liu \etal 2014):
\numparts
\beqn
 \fr{\mathrm{d} \rho}{\mathrm{d} t} = \rho_t+\pu\cdot\nabla\rho=-\rho\vartheta, \lb{NSeqs-ch2a}\\
 \rho \fr{\mathrm{d} \pu}{\mathrm{d} t} = (\rho\pu)_t+\nabla\cdot(\rho\pu\pu) =-\nabla\Pi-\nabla\times(\mu\po), \lb{NSeqs-ch2b}\\
 \rho T \fr{\mathrm{d} s}{\mathrm{d} t} = \rho Ts_t+\rho T\pu\cdot\nabla s =\Phi+ \na \cdot (\kappa \nabla T),\lb{NSeqs-ch2c}
\eeqn
\endnumparts
and the equation of state is
\beq\lb{state-1}
 p=\rho RT.
\eeq
Here $\rho, p, T, s$ are the density, pressure, temperature and entropy, respectively, $\Pi = p - \mu_\theta \vartheta$ is the viscous modified normal stress, $\vartheta = \na \cdot \pu$ the dilatation, $\mu, \mu_\theta, \kappa$ the dynamic transport coefficients of shear, compressing and heat conduction, respectively, $\Phi$ the viscous dissipation, $R$ the gas constant, subscript $t$ denotes the time derivative, and $\mathrm{d}/\mathrm{d}t$ is the material derivative. Because NS equations \eref{NSeqs-ch2a}, \eref{NSeqs-ch2b}, \eref{NSeqs-ch2c} are nonlinear and have infinite degrees of freedom, it is still impossible to obtain its analytical solutions in the general case. However, since in unbounded external-flow there must be a uniform region, we can properly assert that there must be a linear region adjacent to the uniform fluid. In this region, \eref{NSeqs-ch2a}, \eref{NSeqs-ch2b}, \eref{NSeqs-ch2c} can be linearized, making it hopeful to get some significant results.

Let $\epsilon\ll 1$ be a small dimensionless parameter, and denote
\numparts
\beqn
s = c_p (\ep s'+\cdots), \ & &
\rho = \rho_0(1+\ep \rho' +\cdots), \lb{dist-1}\\
p = p_0(1 +\ep p' + \cdots), \ & &
T = T_0(1+\ep T'+\cdots), \lb{dist-2} \\
\pu = \ep \pu' +\cdots, \ & &
\mu = \mu_0(1+\epsilon \mu'+\cdots), \lb{dist-3} \\
\mu_\theta = \mu_{\theta 0}(1+\epsilon \mu'_\theta +\cdots), \ & &
\ka = \ka_0(1+\epsilon \ka' + \cdots), \lb{dist-4}
\eeqn
\endnumparts
where $c_p$ is the specific heat at constant pressure, subscript $0$ refers to the uniform constant value at infinity, and prime $'$ denotes disturbance quantity, which is $O(1)$. Substitute \eref{dist-1}, \eref{dist-2}, \eref{dist-3}, \eref{dist-4} into \eref{NSeqs-ch2a}, \eref{NSeqs-ch2b}, \eref{NSeqs-ch2c} and \eref{state-1}, there is
\numparts
\beqn
 \rho'_t+\nabla\cdot\pu' = m, \lb{NSeqs-1a}\\
 \pu'_t+\frac{a^2}{\gamma} \nabla p' - \nu_\theta \nabla (\nabla\cdot\pu')+\nu\nabla\times(\nabla\times\pu') = \pf, \lb{NSeqs-1b}\\
 s'_t-\alpha\nabla^2T' = Q, \lb{NSeqs-1c}
\eeqn
\endnumparts
and
\beq\lb{state-2}
 p'=\rho'+T'=\gamma(\rho'+s').
\eeq
In the above equations
\beq
a^2=\gamma\frac{p_0}{\rho_0},\quad\alpha=\frac{\kappa}{\rho_0 c_p},\quad\nu_\theta=\frac{\mu_{\theta 0}}{\rho_0},\quad\nu=\frac{\mu_0}{\rho_0}
\eeq
are the speed of sound and kinematic transport coefficients of shear, compressing and heat conduction, respectively, where $\ga$ is the specific heat ratio. To make \eref{NSeqs-1a}, \eref{NSeqs-1b}, \eref{NSeqs-1c} more universal,
three source terms are added, namely, $m, \pf, Q$, which denote sources of mass, momentum, and heat, respectively. They can be regarded as either the remanent nonlinear terms after the linearization of the original NS equations, or the equivalent source terms which represent the contributions of the nonlinear region (including the body) to the linear region. For the former, \eref{NSeqs-1a}, \eref{NSeqs-1b}, \eref{NSeqs-1c} are strictly valid in the entire flow region; while for the latter, they are valid only in the linear flow region.
In this paper, we call equations \eref{NSeqs-1a}, \eref{NSeqs-1b}, \eref{NSeqs-1c} and \eref{state-2} the \textit{linearized NS equations}.

\subsection{Fundamental solutions and decay rate of the decoupled fields}\label{subsec.decoupled}

Before exploring the fundamental solution of the linearized NS equations, we first consider a much simpler case where the flow field can be completely decoupled into a longitudinal field and a transversal field.

Assume all sources are absent ($m=Q=0, \pf=\mathbf{0}$) and introduce the Helmholtz decomposition \eref{u-lt-1} but with $\pu$ replaced by $\pu'$. Then \eref{NSeqs-1a}, \eref{NSeqs-1b}, \eref{NSeqs-1c} can be completely split into a transverse field (Lagerstrom \etal 1949)
\beq\lb{NSeqs-1-T}
(\partial_t-\nu\nabla^2)\pu_\psi=\mathbf{0},
\eeq
and a longitudinal field
\numparts
\beqn
 \rho'_t + \nabla^2\phi = 0,\lb{NSeqs-1-La}\\
 (\partial_t-\nu_\theta\nabla^2)\phi =-\frac{a^2}{\gamma}p', \lb{NSeqs-1-Lb}\\
 s'_t-\alpha\nabla^2 T' = 0. \lb{NSeqs-1-Lc}
\eeqn
\endnumparts
Eliminate the thermodynamic variables in \eref{NSeqs-1-La}, \eref{NSeqs-1-Lb}, \eref{NSeqs-1-Lc}, we obtain
\beq\lb{NSeqs-1-L-1}
 a^2 (\partial_t-\alpha\nabla^2) \nabla^2\phi = (\partial_t-\nu_\theta\nabla^2) (\partial_t-\gamma\alpha\nabla^2) \partial_t\phi.
\eeq

If we further assume (Mao \etal 2010)
\beq\lb{al-nu}
\alpha, \nu_\theta=O(\delta), \quad \epsilon \ll \delta \ll 1.
\eeq
Then from \eref{state-2}, \eref{NSeqs-1-Lb} and \eref{NSeqs-1-Lc} there is
\beq\lb{s-phi2}
 a^2 s' = -(\gamma-1) \alpha \nabla^2\phi + O(\delta^2).
\eeq
In addition, using the state equation \eref{state-2}, \eref{NSeqs-1b} can be rewritten as
\beq\lb{phi-rho-s}
 (\partial_t-\nu_\theta\nabla^2) \phi = -a^2(\rho'+s').
\eeq
Take the time derivative of \eref{phi-rho-s} and use \eref{NSeqs-1-La}, there is
\beq\lb{phi-s}
 (\partial_t^2 -a^2\nabla^2) \phi = \nu_\theta \nabla^2 \partial_t\phi -a^2 \partial_t s'.
\eeq
Finally, by substituting \eref{s-phi2} into \eref{phi-s}, we obtain
\beq\lb{NS-L-2}
 (\partial_t^2-a^2\nabla^2)\phi = b\nabla^2\partial_t\phi,
\eeq
where
\beq
 b\equiv(\gamma-1)\alpha+\nu_\theta
\eeq
is the sound diffusion coefficient (Lighthill 1956).

Evidently, both the decoupled equations \eref{NSeqs-1-T} and \eref{NS-L-2} are of parabolic type,
making the exponential decay rate possible. Specifically, \eref{NSeqs-1-T} is a standard second-order parabolic partial differential equation (PDE), which describes the processes that behave like heat diffusion through a solid and is valid for any transversal variables. In contrast, \eref{NS-L-2} is a third-order PDE of parabolic type, which is also valid for $\phi$, $\vartheta$ and $s'$.
The same equation of \eref{NS-L-2} was obtained by Lagerstrom \etal (1949) without considering the heat transfer so that $b = \nu_\theta$. Later, Wu (1956) also obtained the same equation with heat transfer included but under assumption $Pr_\theta \equiv \nu_\theta /\al =1$,
which is very close to the value of ordinary gases. With these facts we conclude that, if $Q$ is negligible small, which is the usual case as we are considering, then \eref{NSeqs-1c} of the generalized Stokes equations can be omitted and its effect can be represented by replacing $\nu_\theta$ by $b$ in \eref{NSeqs-1b} (see also Mao \etal 2010). Thus, the fundamental solution of the linearized NS equations without heat transfer given by Lagerstrom \etal (1949) can be directly applied to explore the far-field asymptotics of unsteady compressible and viscous external-flow.

General speaking, the fundamental solutions of the decoupled fields can be obtained by integral transforms. Denote the Laplace transform of any function $f(\px,t)$ as $\widetilde{f} (\px, \sigma)$,
such that
\beq
 \widetilde{f} (\px,\sigma) = \int_0^\infty e^{-\sigma t} f(\px,t)\mathrm{d}t, \quad
 f(\px,t) = \frac{1}{2\pi i} \int_{-i\infty}^{i\infty} e^{\sigma t} \widetilde{f} (\px, \sigma) \mathrm{d} \sigma.
\eeq
Take the Laplace transform of the transverse equation \eref{NSeqs-1-T}, denote by superscript $\mathrm{d}$ the decoupled solutions, and let $\widetilde G^{\rm d}_{ \sqrt{ k^2/c_\psi } }$ be the corresponding fundamental solution, there is
\begin{equation}\label{NS-T-3}
  \left( c_\psi \na^2 - k^2 \right) \widetilde G^{\rm d}_{ \sqrt{ \fr{k^2}{c_\psi} } } = 0, \quad
  c_\psi = \nu, \quad k^2 = \sigma.
\end{equation}
Similarly, for the longitudinal field \eref{NS-L-2} there is
\begin{equation}\label{NS-L-3}
  \left( c_\phi \na^2 - k^2 \right) \widetilde G^{\rm d}_{ \sqrt{ \fr{ k^2}{c_\phi} } } =  0, \quad
  c_\phi = b + \fr{a^2}{\sigma}, \quad k^2 = \sigma.
\end{equation}
Let $\be$ represents either $\sqrt{k^2/c_\psi}$ or $\sqrt{k^2/c_\phi}$, the fundamental solution of \eref{NS-T-3} or \eref{NS-L-3} is
\begin{equation}\label{G-T-1}
 \widetilde G^{\rm d}_{ \be } = \left\{
 \begin{array}{l l l}
 \fr{1}{2\be } e^{-\be r},  & r=\sqrt{x^2}, & n=1, \\
 \fr{1}{2\pi} K_0(\be r),  & r=\sqrt{x^2+z^2}, & n=2, \\
 \fr{1}{4 \pi r} e^{-\be r},  & r=\sqrt{x^2+y^2+z^2}, & n=3,
 \end{array}
\right.
\end{equation}
where $K_0$ is the modified Bessel function of the first kind,
\begin{equation}\label{K0}
  K_0( \eta ) \approx - \ln \left( \fr{\eta}{2} \right) - \gamma, \ \eta \rightarrow 0; \quad
  K_0( \eta ) \approx \sqrt{\fr{\pi}{2 \eta }} e^{- \eta }, \ \eta \rightarrow \infty.
\end{equation}
Here (and only here) $\ga = 0.5772 \cdots$ denotes the Euler constant. Note that solutions of lower dimensions can be obtained by the superposition of an infinite number of those of higher dimensions.

We now transform \eref{G-T-1} back to physical space. Denoting $G^{\rm d}_\psi$ as the fundamental solution of the transverse field \eref{NSeqs-1-T}, there is
\begin{eqnarray}\label{G-T-2-3d}
  G^{\rm d}_\psi = \fr{\nu}{ (4 \pi \nu t)^{ \fr{n}{2} }} \exp \left( - \fr{ r^2 }{ 4 \nu t } \right), \quad n = 1, 2, 3.
\end{eqnarray}
Thus, the far-field behaviour of the transverse field is completely determined by the viscous diffusion process. In particular, since all transverse variables, such as $\ppsi,\pu_\psi,\po$, etc.,  satisfy the same equation \eref{NSeqs-1-T} and the source terms must be compact for any observer standing far away enough, any transversal quantity must be exponentially small in the far field.

Similarly, denote $G^{\rm d}_\phi$ the fundamental solution of the longitudinal equation \eref{NS-L-3}.
Then for $n=3$ there is
\begin{eqnarray}\label{G-L-2}
  G^{\rm d}_\phi = \fr{1}{4\pi r} \fr{1}{2\pi i} \int_{-i\infty}^{i\infty} \exp \left( \sigma t - \fr{ \sigma }{ \sqrt{ a^2 + b \sigma } } r \right) \mathrm{d}\sigma.
\end{eqnarray}
By constructing proper integral contour, \eref{G-L-2} reduces to (see \ref{appendix:Laplace} or Lagerstrom \etal 1949)
\begin{eqnarray}\label{G-L-3}
  G^{\rm d}_\phi = \fr{1}{4\pi r} \fr{a}{\sqrt{ 2 \pi b t} } \exp \left\{ - \fr{ (r-a t)^2 }{ 2 b t } \right\}, \quad \fr{a^2 t}{b} \gg 1, \quad n=3.
\end{eqnarray}
This is the viscous fundamental solution of \eref{NS-L-2} for $n=3$, which represents the longitudinal disturbance caused by the initial pulse of unit strength. Away from the wavefront the disturbance decays exponentially, which can only emerge under the \textit{joint dynamic action of unsteadiness, viscosity, and compressibility}. The spherical attenuation factor $1/4\pi r$ in \eref{G-L-3} is a result of the kinematics and independent on the viscosity and the speed of sound. If we neglect this factor,
then for the observer following the wavefront, the  dynamic decaying of longitudinal process will in essence be the same as that of the one-dimensional transverse process, both belonging to the \textit{viscous diffusion mechanism}. Thus, when a shock wave degenerates to a Mach wave or sound wave in the linear far field, its characteristic thickness will increase as time via $\sqrt{b t}$. For $n=1$ there is
\begin{eqnarray}\label{G-L-3-1D}
  G^{\rm d}_\phi = \fr{a}{4} {\rm erfc} \left( \fr{ r-a t}{\sqrt{ 2 b t } } \right), \quad \fr{a^2 t}{b} \gg 1, \quad n=1,
\end{eqnarray}
where ${\rm erfc}$ is the complementary error function,
\begin{eqnarray}\label{erfc}
  {\rm erfc\,} \eta = 2 H(-\eta) + \left[ \fr{1}{\eta} - \fr{1}{2 \eta^3} + O(\eta^{-4}) \right] \fr{e^{-\eta^2}}{\sqrt \pi}, \quad \eta \rightarrow \pm\infty,
\end{eqnarray}
and $H$ is Heaviside function. The 2D fundamental solution can only be written in integral form,
\begin{eqnarray}\label{G-L-2-2}
  G^{\rm d}_\phi = \fr{1}{2\pi} \fr{1}{2\pi i} \int_{-i\infty}^{i\infty} e^{\sigma t} K_0 \left( \fr{ r\sigma }{ \sqrt{ a^2 + b \sigma } } \right) \mathrm{d}\sigma, \quad n=2.
\end{eqnarray}
Nevertheless, due to \eref{K0}, \eref{G-L-2-2} must also decay exponentially in the far field.

In summary, the fundamental solutions of the decoupled longitudinal and transverse fields both decay exponentially in the far field, which confirms our previous assertion that the result obtained by the kinematic method is only an upper bound.

\subsection{Fundamental solutions and decay rate of the coupled fields}\label{subsec.coupled}

We now proceed to investigate the fundamental solution of \eref{NSeqs-1a}, \eref{NSeqs-1b}, \eref{NSeqs-1c} and \eref{state-2}. We can properly assume $Q=0$, namely there is no external heat input (e.g. vaporization or combustion) and the surface of the object is adiabatic. Then, as remarked before, \eref{NSeqs-1c} can be omitted as long as $\nu_\theta$ is replaced by $b$. Further, we only consider the rigid-body case so that $m=0$ as a primary approximation. Thus, \eref{NSeqs-1a}, \eref{NSeqs-1b}, \eref{NSeqs-1c} reduces to
\numparts
\beqn
  \na \cdot \pu' + \pat_t \rho' = 0,  \label{final-2} \\
  \left( b \bT_\phi - \nu \bT_\psi -\pat_t \bI \right) \cdot \pu' - a^2 \na \rho' = -\pf, \label{final-1}
\eeqn
\endnumparts
where
\begin{eqnarray}\label{T12}
  \bT_\phi \equiv \na \na, \quad
  \bT_\psi \equiv \na \na-\na^2 \bI,
\end{eqnarray}
are two linear differential matrix operators, and $\bI$ is the unit matrix. Searching for the dynamic far-field velocity now amounts to finding the fundamental solution of \eref{final-2}, \eref{final-1}.

Similar to the steady case, here the source term $\pf$ is also singular when viewed in the linear far field. However, now it can move arbitrarily in the space, and hence $\de(\px)$ in \eref{f-st} should be replaced by $\delta \left(\px - \int_0^t \pu_B {\rm d}t \right)$, where $\pu_B$ is the velocity of the body. Moreover, the impulse strength or total force $\pF$ alone in \eref{f-st} is insufficient for describing the strength of $\pf$ in unsteady case of \eref{final-1} which, at least, should explicitly include the effect of the body motion. A corresponding strength, say $\widehat \pF$, must only be time-dependent, i.e., $\widehat \pF = \widehat \pF(t)$, and should be directly related to the total force $\pF$ and the body motion. Here, to determine $\widehat \pF$ an additional condition is required.
For example, from the momentum leakage paradox of incompressible flow (Wu 1981, Saffman 1992),
there is
\begin{equation}\label{f-in2}
  \widehat \pF = -\fr{1}{\rho_0} \pF + \fr{\rm d}{ {\textrm d} t} \int_B \pu {\textrm d} V
\end{equation}
or
\begin{equation}\label{f-in}
  \pf(\px,t) = \left( -\fr{1}{\rho_0} \pF + \fr{\rm d}{ {\textrm d} t} \int_B \pu {\textrm d} V \right) \delta \left(\px - \int_0^t \pu_B {\rm d}t \right).
\end{equation}
Note that \eref{f-in2} is exactly the dipole strength used in the study of sound waves (Lighthill 1978). But for compressible flow the expression of $\widehat \pF$ is not clear yet. Nevertheless, it is certain that the source term is singular in space. Consequently, it seems to be impossible to make the Helmholtz decomposition of $\pf$. This is consistent with the fact that even if the entire disturbance flow field is sufficiently weak and can be linearized, the boundary coupling between the longitudinal and transversal processes still exists for viscous flow (Wu \etal 2006, Liu \etal 2015). Thus, in contrast to the preceding subsection, here we have to consider the fundamental solution of the coupled equations \eref{final-2}, \eref{final-1}.

Now assume that the body starts moving at $t=0$ such that we can apply the Laplace transform to \eref{final-2}, \eref{final-1}. After eliminating the density term $\widetilde \rho$, there is
\begin{equation}\label{Stokes-1}
  \left( c_\phi \bT_\phi - c_\psi \bT_\psi - k^2 \bI \right) \cdot \widetilde \pu = - \widetilde \pf.
\end{equation}
Denote $\widetilde \bG$ as the fundamental solution of \eref{Stokes-1}, which can be obtained by the theorem (Lagerstrom \etal 1949):

\textbf{Theorem}
\textit{If $\bT_\phi$ and $\bT_\psi$ are linear differential matrix operators such that
\begin{equation}\label{d43}
  \bT_\phi \cdot \bT_\psi = \bT_\psi \cdot \bT_\phi = {\bf 0}, \quad
  \bT_\phi - \bT_\psi = L \bI,
\end{equation}
where $\bI$ is the $n$-dimensional unit matrix, $L$ is a scalar linear differential operator,
then the fundamental solution $\widetilde \bG(\px,\pxi)$ of \eref{Stokes-1} is given by
\begin{equation}\lb{d90}
  \widetilde \bG(\px,\pxi) = \fr{1}{k^2} \left( \bT_\phi \widetilde G_{\sqrt{\fr{k^2}{c_\phi}}} - \bT_\psi \widetilde G_{\sqrt{\fr{k^2}{c_\psi}}} \right),
\end{equation}
where $\widetilde G_\be(x,\xi)$ is the fundamental solution of the scalar differential operator $L-\be^2$.}

In particular, from \eref{T12} we know that $\widetilde G_\beta = \widetilde G_\beta^{\rm d}$ is still given by \eref{G-T-1}. Thus, the solution of \eref{Stokes-1} is
\begin{equation}\label{u-G}
 \widetilde \pu(\px, \sigma) = \int \widetilde \bG(\px-\pxi, \sigma) \cdot \widetilde \pf(\pxi, \sigma) \mathrm{d}\pxi.
\end{equation}
Transforming back to physical space, we finally obtain the far-field velocity expression
\begin{equation}\label{u-G}
 \pu(\px, t) = \int \bG(\px-\pxi, t-\tau) \cdot \pf(\pxi, \tau) \mathrm{d} \pxi \mathrm{d} \tau,
\end{equation}
where $\bG(r,t)$ is the inverse Laplace transform of $\widetilde \bG(r,\sigma)$,
\begin{equation}\label{G}
  \bG = \bT_\phi \left\{ \fr{1}{2\pi i} \dashint_{-i\infty}^{i\infty} e^{\sigma t} \widetilde G_{\sqrt{ \fr{k^2}{c_\phi}}} \fr{\mathrm{d}\sigma}{k^2} \right\}
  - \bT_\psi \left\{ \fr{1}{2\pi i} \dashint_{-i\infty}^{i\infty} e^{\sigma t} \widetilde G_{\sqrt{ \fr{k^2}{c_\psi}}} \fr{\mathrm{d}\sigma}{k^2}  \right\}.
\end{equation}
Note that the finite-part integral operator $\dashint$ has been adopted since the integrals in \eref{G} are usually divergent.

Because in unsteady compressible and viscous flow the source term $\pf$ can always be regarded as compact, to obtain the far-field asymptotics of unbounded external-flow we can only focus on the behavior of $\bG$ in the far field. In particular, from \eref{G} the fundamental solution $\bG$ can also be decomposed into a longitudinal part and a transverse part,
\begin{equation}\label{d64}
 \bG(r,t) = \bG_\phi(r,t) + \bG_\psi(r,t),
\end{equation}
where
\begin{eqnarray}
  \bG_\phi = \bT_\phi G_\phi, \quad
  \bG_\psi = -\bT_\psi G_\psi,
\end{eqnarray}
and
\begin{equation}\label{G-1}
  G_\al \equiv \fr{1}{2\pi i} \dashint_{-i\infty}^{i\infty} e^{\sigma t} \widetilde G_{\sqrt{ \fr{\sigma}{c_\al}}} \fr{\mathrm{d}\sigma}{\sigma}, \quad \al = \phi, \psi.
\end{equation}

Consider the transversal field first. Then for $n=3$ there is (Lagerstrom \etal 1949)
\begin{equation}\label{G-1}
  G_\psi = \fr{1}{4 \pi r} \fr{1}{2\pi i} \int_{-i\infty}^{i\infty} \exp \left( \sigma t - \sqrt{ \fr{\sigma}{\nu} } r \right)  \fr{\mathrm{d}\sigma}{\sigma} =\fr{1}{4\pi r} {\rm erfc\,} \left( \fr{r}{2\sqrt{\nu t}} \right).
\end{equation}
From \eref{erfc} we see that when $r / 2 \sqrt{\nu t} \gg 1$ all transverse quantities ($\pu_\psi$, $\po$, etc.) decay exponentially. This is also true for the case $n=2$. In particular, there is (Lagerstrom \etal 1949)
\begin{equation}\label{G-1-2}
  G_\psi = \fr{1}{2 \pi} \fr{1}{2\pi i} \dashint_{-i\infty}^{i\infty} e^{\sigma t} K_0 \left( \sqrt{ \fr{\sigma}{\nu} } r \right)  \fr{\mathrm{d}\sigma}{\sigma} = \fr{1}{4 \pi} {\rm E_1\,} \left( \fr{r^2}{4 \nu t} \right),
\end{equation}
where ${\rm E_1\,}$ is the exponential integral function, which satisfies
\begin{eqnarray}\label{E-1}
  {\rm E_1\,} (\eta) =
  \left[ \fr{1}{\eta} + O \left( \eta^{-2} \right) \right] e^{-\eta}, \quad \eta \rightarrow \infty.
\end{eqnarray}

Next, consider the longitudinal field. Then for $n=3$ there is (see \ref{appendix:Laplace} or  Lagerstrom \etal 1949)
\begin{equation}\label{G-1}
  G_\phi = \fr{1}{4 \pi r} \fr{1}{2\pi i} \int_{-i\infty}^{i\infty} \exp \left( \sigma t - \fr{ \sigma r}{ \sqrt{ a^2 + b \sigma } } \right)  \fr{\mathrm{d}\sigma}{\sigma} = \fr{1}{8 \pi r} {\rm erfc\,} \left( \fr{r-at}{\sqrt{2 b t}} \right).
\end{equation}
From \eref{erfc} we see that, for the true infinity where $(r-at)/\sqrt{2 b t} \gg 1$, all longitudinal quantities decay exponentially. For $n=2$, however, $G_\phi$ can only be expressed by contour integral,
\begin{equation}\label{G-2}
  G_\phi = \fr{1}{2\pi} \fr{1}{2\pi i} \dashint_{-i\infty}^{i\infty} e^{\sigma t} K_0 \left( \fr{\sigma r}{\sqrt{ a^2+b\sigma}} \right) \fr{\mathrm{d}\sigma}{\sigma},
\end{equation}
but, due to \eref{K0}, the asymptotic behaviour of exponential decaying is still ensured.

In contrast to the above decay rate at true infinity for compressible flow, for incompressible flow we can only set $(at-r)/\sqrt{2 b t} \gg 1$ and hence \eref{G-1} reduces to $G_\phi = 1 / 4 \pi r$, while in 2D \eref{G-2} reduces to $G_\phi = - (1/2\pi) \ln r$ due to \eref{K0}. Thus, we have explicitly revealed that the far-field asymptotics obtained by Poisson's equation or kinematic method are not the truly far-field asymptotics of $V_\infty$ but only a subspace of $V_\infty$, namely the incompressible flow zone $V_{\rm inc}$ . Moreover, since the transversal field decays exponentially as $r/2 \sqrt{\nu t} \rightarrow \infty$, in the study of the far field of incompressible flow, we only need to consider the longitudinal field, where
\begin{equation}\label{G-L-in}
 G_\phi = \left\{
 \begin{array}{cll}
 -\fr{1}{2\pi} \ln r,  & r=\sqrt{x^2+z^2}, & n=2, \\
 \fr{1}{4 \pi r},  & r=\sqrt{x^2+y^2+z^2}, & n=3,
 \end{array}
\right.
\end{equation}
which is the time-independent fundamental solution of the Poisson equation. This observation naturally leads to our finding of zonal structure of unbounded external-flow, which is discussed in the next section.

\section{Zonal structure of unbounded flow domains}\label{sec.zonal}

The decay rates of disturbances in far field estimated in the preceding section should have been optimal, as they are established both kinematically and kinetically. We have seen that the furthest far-field flow adjacent to the uniform fluid at infinity is \textit{unsteady, viscous and compressible}, where all disturbances degenerate to viscous sound waves and damp out exponentially.
This is consistent with the fact that only sound waves can propagate themselves without external forces and thus travel furthest (Lighthill 1978). They are annihilated there not by nonlinear dissipation but by dispersion (Lighthill 1956). At the same time, we have also encountered some concepts of `infinity' and decay rate in various externally unbounded flow models, which are not exponential.
This situation suggests a zonal structure of flow domains used by different theoretical models in near and far field, such as incompressible flow, steady flow, and inviscid flow, ect., of which a thorough clarification as we attempt below may strengthen and deepen our physical understandings on the overall picture of this type of flow. This zonal structure has somewhat analogy with the various wall regions and layers in turbulent boundary layer, where the mean velocities satisfy different approximate rules, say, linear law in viscous sublayer and log-law in log-law region, which are crucial for the high-accurate modeling of turbulent flow. Although how to apply the zonal structure into modern computational fluid dynamics (CFD) is not clear yet, its importance can never be ignored. In addition, with the help of the zonal structure, we can easily discriminate the advantages and disadvantages of various aerodynamic theories and clarify some paradoxes concerning far-field behaviour. This is done in the next section. In this section, the zonal structure is illustrated first.

As before, the externally unbounded flow is assumed to be caused by a finite body moving through it, and the flow plus body fulfills the entire free space $V_\infty$ with fluid at rest or in uniform status at infinity. We add that the body and fluid are initially at rest at $t\leq 0$, and let the airfoil start moving at $t=0$, reaching its final state $\pu_B = - U \pe_x$ at $t=t_s$. Then a preliminary zonal structure is shown in \Fref{fig.flow-field}, which for the sake of illustration only describes the disturbance development caused by the low-speed or subsonic flight of an airfoil. This figure is yet incomplete; some further classification will be introduced below. But since the far-field decay law of viscous, unsteady and compressible flow has been fully clarified, no more discussion on it is needed.

Adjacent to the uniform fluid at infinity, there must be a zone where the disturbance intensity $E$ decays sufficiently small, say $E = O (\ep)$ with $\ep \ll 1$, so that the governing equations can be linearized. We call this zone the linear zone or linear far-field, and denote it by $V_{\rm L}$. This linear zone encloses a nonlinear zone $V_{\rm NL}$. While the disturbance intensity in the latter is $E = O (\ep^0)$, in the former it is $E = O (\ep^m)$, $m>1$. Thus, $V_{\rm L}$ should locate between the nonlinear zone $V_{\rm NL}$ and uniform zone (i.e. the zone between the solid loop and dashed loop in \Fref{fig.flow-field}). Although the existence of this linear far-field is assumed based on physical intuition and a mathematical rigorous proof is still lacking, one can check the existence from the behaviour of obtained analytical solutions of linearized equations. On the other hand, although the details of how the flow transforms from $V_{\rm NL}$ to $V_{\rm L}$ is of great interest, it is beyond the scope of this paper.

\subsection{Nonlinear near field, steady or unsteady}

We start from the most inner zone enclosed by the solid loop in \Fref{fig.flow-field}. Since the viscous fluid has to satisfy the no-slip and no-penetration conditions on a solid wall, thus in the region very close to the wall, such as the viscous sublayer of boundary layer and initial segments of free shear layer, either laminar or turbulent, the flow must be rotational with strong effect of viscosity. Once leaving the body surface, the flow quickly evolve nonlinearly as the characteristic feature of near-field flow. The flow can be either incompressible or compressible, and either intrinsically unsteady, or steady viewed in the reference frame fixed to the body.

Denote by $c$ the length scale of the body, say the chord length of the airfoil, then the nonlinear zone surround the body occupies a volume $V_{\rm NL} = O(c^n)$. Obviously, in $V_{\rm NL}$ the flow exhibits its full complexity, in particular at large Reynolds numbers and Mach numbers. The governing equation in $V_{\rm NL}$ is the fully nonlinear NS equations of infinity degrees of freedom, and it is almost impossible to find its analytical or asymptotic solution in general case. This is the major place where CFD and advanced experimental techniques show their full power in revealing the detailed complex flow structures and processes. In the computation, the flow conditions at the external boundary of computational domain have to be prescribed, which depends on what far-field zone it will be right outside the domain.

\subsection{Steady far field}

\begin{figure}
  \centering
  \includegraphics[width=0.7\textwidth]{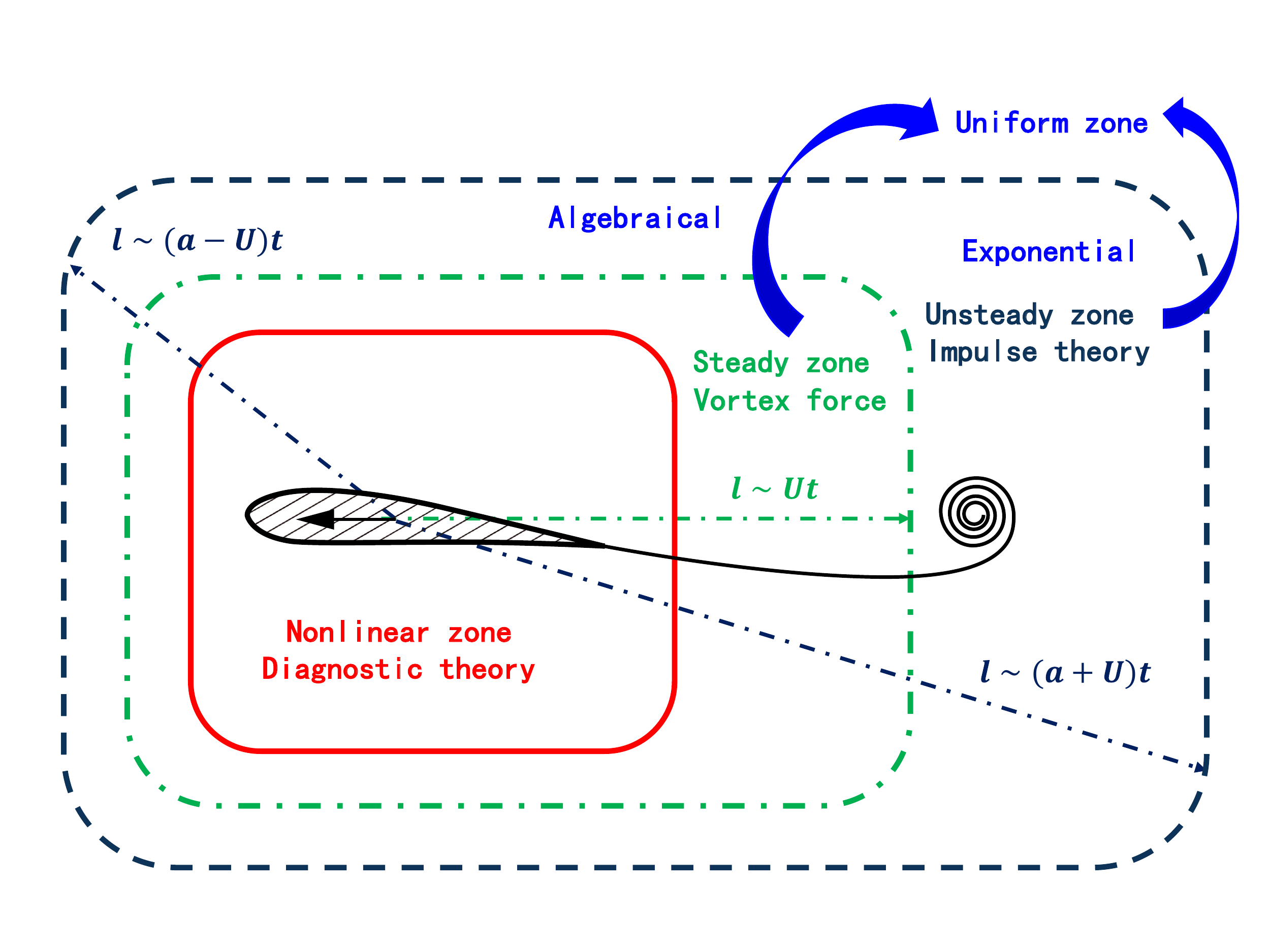}
  \caption{Sketch of zonal structure of unbounded external-flow}
  \label{fig.flow-field}
\end{figure}

Return to the airfoil motion in \Fref{fig.flow-field}. The transient dynamic process and causal mechanisms from the airfoil starts motion to the establishment of its lift, along with a starting vortex shedding downstream to ensure total-circulation conservation, has been addressed in detail by Zhu \etal (2015) and references therein, so our concern here is only the zonal structures of the flow field after the airfoil has reached or almost reached its final state.

As time goes on, the starting vortex continually moves away to sufficiently large distance behind the airfoil, such that its effect on the flow field near the airfoil is negligible. Actually, such a distance needs not to be very large; for the starting problem of 2D thin airfoil, von K\'arm\'an and Sears (1938) have theoretically proven that a few chord lengths will be enough. Therefore, if we shift the reference frame from the one fixed to the still fluid at infinity to that fixed to the airfoil, when this distance satisfies the condition $l \sim Ut \gg c$ (see \Fref{fig.flow-field}), there can be a zone which excludes the starting vortex and in which the flow is steady or statistical steady, with uniform incoming flow velocity $\pU =U\pe_x$. We call such a zone the steady zone and denote it by $V_{\rm st}$ (i.e. the zone enclosed by the dashed-dotted loop in \Fref{fig.flow-field}). In subsonic flow, when $t \ge t_s$, although the total amount of vorticity shedding off the airfoil is zero (Liu \etal 2015, Liu 2016), there is always vorticity with the same magnitude but different signs shedding off the upper and lower airfoil surfaces, respectively. This region where the vorticity itself is nonzero but its total flux is zero is called the steady wake, which connects the starting vortex and airfoil's boundary layers.

The above overall picture will have some modification for transonic or supersonic incoming flow, where the specific near-field flow structures can be very complicated due to the appearance of shock waves not shown in \Fref{fig.flow-field}, which can make the total vorticity flux no longer be zero (Liu \etal 2015). However, this complexity does not change the corresponding zonal structures, provided that the fluid is viscous.

Owing to the fact that part of vortical wake must be inevitably excluded from $V_{\rm st}$ in both incompressible and compressible flows, the transverse Oseen equation indicates that as $r\rightarrow \infty$ in $V_{\rm st}$ the velocity can only decay algebraically (Liu \etal 2015, Liu 2016).
Evidently, $V_{\rm st}$ must be a true subspace of $V_\infty$.

\subsection{Unsteady far field, incompressible}

If one needs to investigate the above airfoil flow in an exceedingly large region that encloses the starting vortex, the whole disturbance zone $V_\infty$ enclosed by the dashed loop in \Fref{fig.flow-field} has to be taken into consideration, and then the flow must be intrinsically unsteady. Wu (1981, 2005) was one of the first to emphasize this relationship between steady and unsteady flows by thorough physical discussion. But this issue has not yet become trivial. More awareness of and attention to it are needed in aerodynamics community.

Unsteady incompressible far field, however, is not yet able to enjoy exponential decay. This has been explicitly shown by \eref{G-L-in}, because the longitudinal equation \eref{NS-L-2} degenerates to a Laplace equation for $\phi$. In fact, incompressibility assumption is incompatible with the furthest zone at far field. Specifically, as stressed by Landau \& Lifshitz (1987), for steady flow to be regarded as incompressible, the familiar condition $|\pu| \ll a$ is sufficient; but for unsteady flow, a further condition has to be added: if $t$ and $l$ are the temporal and spacial scales over which the flow undergoes significant changes, then there should be $t \gg l/a$. Now the first condition can always be satisfied as along as $\pu_B$ keeps small enough. But the second one cannot as the truly far-field asymptotics $r \rightarrow \infty$ is reached where $l \sim V^{1/n}_\infty$, although it can at near and middle fields. Therefore, the incompressible flow zone, say $V_{\rm inc}$, is also a true subspace of $V_\infty$.

Here we recall a classic paradox that the total momentum of unbounded incompressible fluid has only conditional convergence, and the total angular momentum diverges. Evidently, the paradox will disappear at once as one realizes the incompressibility domain cannot reach the true infinity, but is surrounded by a viscous, compressible, unsteady and linear zone with exponential decay. Furthermore, ignoring this fact has also caused the well-known paradox of the same root as the poor behaviour of total momentum and angular momentum for incompressible flow: there must be $\pI/n$ ($\pI$ is the impulse, see \Eref{impulse} below) portion of total momentum escaping out of a spherical domain of arbitrarily large radius, although the flow there is irrotational. The paradox was removed by Landau \& Lifshitz (1987), who pointed out that the far-field flow is compressible and the escaped momentum is carried away by sound wave, see also Saffman (1992).

Actually, unsteady near-field incompressible flow surrounded by compressible far field has been a well-known and effective model in the field of aeroacoustics ever since Lighthill (1952) constructed the acoustic analogy theory. In that theory the source of sound (say an unsteady vorticity field) can be treated incompressible in a compact region, which emits sound as very weak disturbance waves to far field but is not affected by the waves.

\subsection{Role of viscosity}\label{subsec.viscosity}

As said before, very near the body surface the fluid  viscosity plays a key role for the satisfaction of the adherence condition and the motion inside the boundary layer. At large Reynolds number $Re$, the explicit viscous effect outside the strong shear layers can often be neglected. As a common concept, the viscosity can then be neglected all the way till the far field. But once again this concept is incorrect. Without viscosity the far-field sound waves cannot be annihilated to ensure the smooth exponential transition to the uniform fluid at infinity. More specifically, Liu \etal (2015) have shown that,  although in subsonic regime the leading-order far-field behaviour of the flow is still of inviscid nature, in transonic and supersonic flow regimes {\it no} linear far field can exist without viscosity. Their numerical simulation has confirmed the analytically obtained  asymptotic behaviour and location of the {\it viscous} linear far field in high-speed flow regime.

\section{Relevance to external-flow aerodynamics}\label{sec.force}

The zonal structure of externally unbounded flow bears close relevance to aerodynamics, of which the central concern is the force and moment acted to the moving body by the fluid. In this section we examine how the zonal structure influences aerodynamic problems formulated by various flow models. Once again the viscous and compressible unsteady flow does not need to be discussed; it is just a perfect model with which other flow models are to be compared.

Aerodynamic theory has been developed along two approaches, far-field and near-field, in a combined manner. The former uses linearized far-field equations and can obtain concise, accurate and universal force formulas, but leaving the determination of the value of the key variable thereof to the latter for specific problems. For example, the famous Kutta-Joukowski (KJ) theorem $L=\rho U\Ga$, the first cornerstone of modern aerodynamics, was first derived by Joukowski (1906) using steady far-field approach (for its version in modern language see Batchelor 1967, pp.~406-407). But the value of circulation $\Ga$ has to be fixed by near-field theory under prescribed specific body geometry and flow condition, including the Kutta condition for inviscid flow model. Conversely, a nonlinear near-field theory can check the far-field results by taking its leading-order approximation as $r\rightarrow \infty$. Below we organize our discussion based on this classification of far- and near-field approaches and their combination.

\subsection{Near-field low-speed aerodynamics}

Among classic low-speed aerodynamic theories developed before computer era, we consider the vortex-force theory for steady flow and impulse theory for unsteady flow as two most brilliant pearls due to their neatness in form, physical insight implied thereby, and generality in their respective zones. Here we use these theories to demonstrate the importance of identifying the proper flow zones to which different aerodynamics theories can apply.

We remark that both these theories belong to near-field type effective in nonlinear zone $V_{\rm NL}$, since they involve inherently nonlinear domain integrals. Fortunately, due to the rapid development of CFD it has become a routine task to solve the NS equation numerically. Thus, the main task of corresponding modern aerodynamics theory should be switched from finding analytical solutions to identifying the key quantitative contribution of specific flow structures and dynamic processes to the aerodynamic performance. We call such a theory the \textit{diagnosis theory of complex flows}. Namely, except a few special cases where the flow is fully attached and force formulas can be simplified by small perturbation techniques, the power of these theories could only be fully appreciated when they are used to diagnose detailed flow-field data already obtained experimentally or numerically.

In the steady zone $V_{\rm st}$ as shown in \Fref{fig.flow-field}, the most significant low-speed flow structure is the boundary layer and vortical wake. Accordingly, the most beautiful aerodynamic theory is the {\it vortex-force theory} (Prandtl 1918, Saffman 1992), which was originally formulated for inviscid flow or at the limit $Re\rightarrow \infty$. In this theory, the force and moment are expressed by the integrals of the Lamb vector $\po\times\pu$ and its moment $\px\times (\po\times \pu)$, respectively. For example, with $\rho_0$ being constant reference density, the force reads
\beq\lb{VF}
\pF = -\rho_0\int_{V_{\rm st}}\po\times \pu \mathrm{d}V,
\eeq
which not only has excellent convergence property and clear asymptotic form as $r\rightarrow \infty$,
but also can reveal the specific contributions of the flow structures to the force. In particular, in 2D \Eref{VF} degenerates to the KJ formula $L=\rho U\Ga$ (Prandtl 1918, von K\'arm\'an \& Burgers 1935), while in 3D it contains reduced drag with its linearized approximation leading to Prandtl's lifting-line theory. Now the vortex-force theory has been generalized to viscous and unsteady flow, by adding a wake-plane integral to account for the form drag and a domain integral of $\px\times \po_{t}$ to account for the flow unsteadiness (Wu \etal 2007, Wu \etal 2015).

On the other hand, in the unsteady zone $V_\infty$ as shown in \Fref{fig.flow-field}, the most significant flow structure in low-speed aerodynamics is also the boundary layer and vortical wake, but now the unsteady motion of the starting vortex is included. In this case the most beautiful aerodynamic theory is the {\it impulse theory} or {\it vorticity-moment theory} (Burgers 1920, Wu 1981, Lighthill 1986), in which the force and moment are expressed by the time rate of the integrals of $\px\times \po/(n-1)$ and $r^2\po/2$ for $n=2,3$, respectively. For example, For fluid occupying $V_f$, which extends to infinity and bounded internally by body surface $\pat B$, the total force reads (Wu 1981)
\beq\lb{impulse}
\pF = -\rho_0\fr{\mathrm{d}\pI}{\mathrm{d}t}+\rho_0\fr{\mathrm{d}}{\mathrm{d}t}\int_B \pu_B \mathrm{d}V, \ \ \ \pI = \fr{1}{n-1}\int_{V_\infty}\px\times \po \mathrm{d}V.
\eeq
The nonlinearity and kinetic content of the theory will show up once the time-rate operator $\mathrm{d}/\mathrm{d}t$ is shifted into the integral (Saffman 1992, Wu \etal 2015). Since vorticity is physically compact, the impulse theory is very suitable for the forces acted on bodies that have arbitrary motion and deformation of the body at any Reynolds numbers. In particular, it has now been the primary choice in the force analysis of animal locomotion; for a recent example see Meng \& Sun (2015).

To fit the need for using the theory in practical experiments and computations, one may shrink the domain boundary in impulse theory to a finite one to explore its full generality. The result is exactly the recovery of the force formula of the generalized vortex-force theory. However, the concept of impulse was introduced as an artifact that is `applied to a limited portion of the fluid in order to generate the whole of the given motion from rest' (Batchelor 1967, p.~518). The impulse $\pI$ does not really equal the total momentum; their difference is an integral, say $\pS$, over the outer boundary of flow domain. The very neat formula \eref{impulse} holds only if the vortex system under study is compact and at its outer boundary the flow is irrotational. It can be proven that once the domain boundary cuts the vortical wake as is inevitable for steady flow, the term $\mathrm{d}\pS/\mathrm{d}t$ will immediately become very complicated and \eref{impulse} no longer holds.
In this case, the artificial splitting of the total momentum into $\pI$ and $\pS$ is physically meaningless. Actually, the best one can do with the finite-domain impulse theory is requiring the vortex system under study is compact so at the boundary the flow remains irrotational, which is fortunately the case for those animal motions that generate a series of nearly compact vortex rings in the wake.

\subsection{Near-field high-speed aerodynamics}

The above remark on near-field low-speed aerodynamic theories on their nonlinearity also applies to near-field high-speed aerodynamics. But the latter is much more complicated than the former, due to longitudinal process caused by compressibility and entropy change associated with shocks. Thus, classic high-speed aerodynamics can only rely on further theoretical models which are oversimplified in two aspects, implying that it has less generality compared to its low-speed counterpart. First, the inviscid-flow assumption is made except in attached boundary layers. Second, the full NS equations are mostly replaced by small disturbance potential-flow equations. In modern aerodynamics, as one's major concern has become complex flows with steady or unsteady flow separation and separated flows with free shear layers, shocks and vortices, the framework of classic high-speed aerodynamics with these oversimplifications has inevitably made it far behind the need of modern numerical and physical experiments.

On the first oversimplification, it has been addressed in \S~\ref{subsec.viscosity} that the neglect of viscosity makes it impossible to construct linear far-field theory, especially for transonic and supersonic flows. As a remedy of this lacking, Cole \& Cook (1986) have to introduce nonlinearity to inviscid transonic `far field'. But the corresponding result is evidently neither smooth nor truly far field. In fact, while modern CFD scheme for high-speed aerodynamics can resolve the flow structure in shear layers at the scale of $O(Re^{-1/2})$, it is not yet able to resolve viscous shock layers at the scale of $O(Re^{-1})$ (laminar flow for example). Thus, at large Reynolds numbers, away from thin boundary layers and vortical wake, the global external flow can indeed be assumed inviscid plus shock discontinuity. However, the viscosity has to be recovered in the far field to ensure the exponential decay of all disturbance quantities, and to construct physically correct far-field theory. The recovery of viscosity in far field but retaining inviscid assumption in `middle field' is quite similar to the recovery of compressibility in far field but retaining incompressible assumption elsewhere in low-speed aerodynamics. A recently developed far-field theory for viscous and compressible steady flow will be highlighted in the next subsection.

To remove the second oversimplification of classic high-speed aerodynamics as well as to include viscous effects, one finds that besides the Lamb vector there is a new longitudinal term, namely the gradient of the density (Chang \etal 1998, Wu \etal 2006, Xu \etal 2010), which can be further split into two parts, namely the Mach-number weighted Lamb vector and the gradient of the Mach-number weighted temperature, of which the latter represents the contribution of the longitudinal process to the force and moment (Liu \etal 2014, Liu 2016). The resulted {\it longitudinal-transverse force theory} matches CFD perfectly as they are based on the same full NS equations.

Following the same tactics of Wu (1981), Huang (1994) has attempted to generalize the impulse theory to compressible flow, by replacing the vorticity by the `dynamic vorticity' $\po^* \equiv \na \times (\rho \pu)$. Not mentioned by Huang is that, the linearized governing equation of $\po^*$ is the same as that of $\po$ multiplied by a constant $\rho_0$ when the frame is fixed on the infinity still fluid. Thus, we can dynamically prove that $\po^*$ has the same compactness as that of $\po$, making the compressible impulse theory be rigorous. Unfortunately, there is still no direct application of this theory to practical diagnosis of complex compressible flows.

\subsection{Steady far-field aerodynamics}

A necessary condition for the construction of complete far-field aerodynamic force theories is that the total force and moment can be solely expressed by control-surface integrals. As remarked in \S~\ref{subsec.kinematics}, this can not be fulfilled for unsteady flow in general case due to the volume integral of $(\rho \pu)_t$. Thus, the far-field force theory is still uncompleted. However, the total forces of steady flow and incompressible flow are two pleasant exceptions, where for the former the unsteady term disappears automatically while for the latter (and force only) it can be transformed to a related boundary-integral (Noca \etal 1999, Wu \etal 2005).

For the flow in $V_{\rm L}$, the governing equation can be reduced to its linearized form.
In specific, when the frame of reference is fixed to still fluid at infinity, the linearized equation is the linearized NS equations \eref{NSeqs-1a}, \eref{NSeqs-1b}, \eref{NSeqs-1c} and \eref{state-2}; while when the frame of reference is fixed on the body, it is the steady/unsteady Oseen equation. Though these two equations are fully equivalent in physics, they have their individual advantages and disadvantages in specific problems.
For example, in steady zone $V_{\rm st}$, the Oseen equation is more convenient (Liu \etal 2015);
while in the unsteady zone, especially when the speed of the object is time-dependent, $\pU=\pU(t)$,
the linearized NS equations are more convenient.

Using the far-field method, Liu et al. (2015) obtained a universal theory for the aerodynamic lift and drag on a body in 2D, steady, viscous and compressible external flow, effective from incompressible all the way to supersonic regimes. Two sets of total-force formulas have been derived.
In the first set, the lift $L$ and drag $D$ are given exactly and universally by the contour integrals of velocity scalar potential $\phi$ and vortical stream-function $\ppsi$, respectively:
\numparts
\beqn\lb{F2D}
\pF =  \rho_0(\pU\times \pGamma_\phi + \pU Q_\psi),\lb{F2Da}\\
\pGamma_\phi \equiv \int_S\pn\times \na \phi \mathrm{d}S,\ \ \ Q_\psi \equiv -\int_S(\pn\times \na)\cdot \ppsi \mathrm{d}S,\lb{F2Db}
\eeqn
\endnumparts
where $S$ is an arbitrary contour enclosing the body, which may even lie in $V_{\rm NL}$, the highly nonlinear zone around the body. This result has the same form as but significantly generalizes the classic KJ inviscid lift formula (Joukowski 1906) and Filon's (1926) viscous drag formula for incompressible flow to any viscous and compressible flow. However, the universality of this set of formulas is at the expense that $\phi$ and $\ppsi$ are not directly testable by experiment or computation. Thus, the second set of formulas was derived in terms of the physically testable quantities only, which is the asymptotic approximation of the first set and holds only in $V_{\rm L}$. The most remarkable feather of this set is that $L$ and $D$ are solely expressed in terms of far-field vorticity integral even if the flow is supersonic. On the $(x,z)$-plane with $\po =(0,\om,0)$ and $\ppsi = (0,\psi,0)$, the result reads
\beq\lb{LD2Dfar}
L = \lim_{r\rightarrow \infty}\int_{V_{\rm st}}\om \mathrm{d}V =[\![\phi]\!], \quad
D =\lim_{r\rightarrow \infty}\int_Wz\om \mathrm{d}z =[\![\psi]\!],
\eeq
where $W$ is a far-field wake plane and $[\![\cdot]\!]$ denotes the jump of potentials due to the double-connectivity of 2D flow domain. Therefore, \textit{no matter how many interacting processes could appear in a nonlinear complex near-field flow, only the vorticity field has the furthest downstream extension as the sole signature of the complex flow in far field, of which the distribution can faithfully capture the total aerodynamic force}. In addition, Liu \etal (2015) have also proven that the far-field velocity of viscous and compressible flow (including subsonic, transonic and supersonic flows) decays algebraically as $r \rightarrow \infty$ in $V_{\rm st}$ (\Fref{fig.flow-field}), and have given a direct evidence of the existence of linear zone, by comparing the theoretical estimates of the location of the linear zone with the results obtained by numerical simulation.

Following the same strategy of Liu \etal (2015), the above result has been extended to 3D by Liu (2016).

\subsection{Unsteady far-field theory}

For an externally unbounded incompressible flow over an arbitrarily moving and deforming body $B$ with prescribed velocity distribution $\pu_B(\px,t)$ at its surface $\pat B$, Wu \etal (2005) have obtained an exact total-force formula solely in terms of boundary integrals over $\pat B$ and a control surface $\Sigma$. It can be written as
\beqn
 \pF &=& -\rho\fr{\mathrm{d}}{\mathrm{d}t} \left[ \int_{\pat B} \px (\pu_B\cdot\pn) \mathrm{d}S +\int_\Sigma \px (\pu\cdot\pn) \mathrm{d}S \right] \nonumber \\
 & & -\fr{\rho}{n-1} \int_\Sigma \px\times (\pn\times \pa) \mathrm{d}S + \int_\Sigma (\rho \ps +\ptau) \mathrm{d}S, \lb{Wu2005}
\eeqn
where $\pa =\mathrm{d}\pu/\mathrm{d}t$ is the material acceleration, $\ps =\pat \po/\pat n$ is the vorticity diffusion flux, and $\ptau =\mu\po\times \pn$ is the shear stress. At far field this formula can be linearized, but similar convergence difficulty as $r\rightarrow \infty$ remains as the total momentum for incompressible flow. Once again, the way out should be the recovery of compressibility at far field.

The 2D version of \eref{Wu2005} was derived independently by Iima (2008) via complex-variable approach. Interestingly, Iima raised a paradox of hovering insects in space: insects maintaining their bodies in a particular position cannot, on average, generate hydrodynamic force if the induced flow is temporally periodic and converges to rest at infinity. Evidently, the same paradox could also true for 3D flow. As a demonstration of how to apply our preceding results, let us resolve this paradox. For convenience we further assume that the insect's body volume can be omitted. Then the momentum source \eref{f-in} reduces to \eref{f-st} but with a time-dependent $\pF(t)$:
\begin{equation}\label{f-ust}
  \pf(\px,t) =-\fr{\de(\px)}{\rho_0}\pF(t).
\end{equation}
Then, by substituting \eref{f-ust} into \eref{u-G} there is
\begin{eqnarray}\label{Stokes}
  \phi = - \fr{1}{\rho_0} \na G_\phi \cdot \int_0^t \pF(\tau) \mathrm{d} \tau.
\end{eqnarray}
It is now clear that although the total force $\pF$ can be assumed to be periodic, the corresponding far-field flow can not. Rather, there must be a constant term in $\pF$ which balances the weight of the body, e.g., $\pF(t) = -m \pgg + \bar{\pF}(t)$, where $m$ is the mass of the insect, $\pgg$ is the gravitational acceleration, and $\bar{\pF}(t)$ is a periodic function whose time-average is zero.
Then, from \eref{Stokes}, there must be a term which is proportional to time, making the far-field flow be non-periodic. In particular, if we omit the term $\bar{\pF}(t)$, which may be small compared to $m \pgg$, then \eref{Stokes} reduces to
\begin{eqnarray}\lb{hovering}
  \phi = \fr{t}{\rho_0} m \pgg \cdot \na G_\phi.
\end{eqnarray}

It should be pointed out that the above argument is based on two conditions. First, since we have omitted the volume of the body or the term $\fr{\rm d}{ {\textrm d} t} \int_B \pu {\textrm d} V$, to ensure $m \pgg$ to be dominant the density of the insect body must be much larger than that of the fluid. In other words, Iima's periodic assumption can be valid for fish swimming. Second, since there will be some terms proportional to $t^2$ in the nonlinear term $\pu\cdot \na \pu$, which for large enough time $t \gg 1$ must prevail other terms in the full NS equations (the viscous term is put aside), to ensure the linear assumption there must be
\begin{equation}\label{linear-c}
  t \ll r^{n+1},
\end{equation}
namely, the nonlinear term is much smaller than other terms.

\section{Conclusions}\label{sec.Conclusion}

This paper studies the asymptotic behaviour of velocity field as $r \rightarrow \infty$, a fundamental issue in all unbounded external-flow problems. Our analysis is based on the assumption that for viscous flow over a finite body, between the innermost nonlinear near field $V_{\rm NL}$ and uniform fluid at infinity there must be a linear field $V_{\rm L}$ where the Navier-Stokes equations can be linearized, leading to a pair of coupled linear, longitudinal and transverse equations. While this assumption is not yet generally proven in mathematic rigour, the existence of linear far field can be checked by physical behaviour of the analytical solutions obtained thereby.

Using this linear far-field analysis both kinematically and kinetically, we found that:

1. The furthest far-field zone adjacent to the uniform fluid at infinity must be unsteady, viscous and compressible, where all disturbances degenerate to viscous sound waves that decay exponentially.
The well-known algebraic decay of velocity field (e.g., Batchelor 1967) is only a kinematic result which, although holds universally, is too conservative and only serves as an upper bound.

2. All flow models simplified from unsteady, viscous and compressible flow, as commonly used in various theoretical and computational studies, fail to satisfy the above exponential decay rule, since they are effective only in certain true subspaces of the free space $V_\infty$. Thus, instead of just assuming the flow to become uniform as $r\rightarrow \infty$ in these models, there is a zonal structure at far field. Specifically, in formulating outer conditions in these models, it should be born in mind that:

--- The steady flow zone $V_{\rm st}$ must be surrounded by an unsteady far field;

--- The incompressible flow zone $V_{\rm inc}$ must be surrounded by a compressible far field; and

--- The inviscid flow zone $V_{\rm inv}$ must be surrounded by a viscous far field.

3. The far-field zonal structure is of close relevance to external-flow aerodynamics. It is demonstrated why aerodynamic theories derived from the above simplified models encounter some difficulties or lead to paradoxes, and how to remove them in terms of the zonal structure.

\ack
This work was partially supported by NSFC (Grant No. 10921202, 11472016) of China.
The authors are very grateful to Profs. Weidong Su and Yipeng Shi, Dr. Jinyang Zhu, Messrs. Shufan Zou and Ankang Gao for their valuable discussions. Special thanks are given to Prof. Zhensu She, who pointed out the analogy between zonal structure in unbounded external-flow and various wall regions in turbulent boundary layer.

\appendix

\section{A kind of inverse Laplace transform}\label{appendix:Laplace}

Consider the following Laplace inverse transform (Lagerstrom \etal 1949):
\beq\label{L-2D}
  L(x, t; a) \equiv  \fr{1}{2\pi i} \int_{-i\infty}^{i\infty} \exp \left( \sigma t - \fr{ \sigma }{ \sqrt{ 1 + \sigma } } x \right) \sigma^{a-1} {\rm d} \sigma,
\eeq
where $t>0$, $x>0$, and $a=0, 1,2, \cdots$.
This integral can be given by the following evaluation
\begin{eqnarray*}
  L(x, t; a) = \fr{1}{2} \fr{\pat^a}{\pat b^a} {\rm erf}\left( \fr{b}{\sqrt{2t}} \right) + \left[ \fr{1}{2} \right]_{a=0} + O \left( t^{- \fr{a+1}{2} } \right), \quad b=t-x,
\end{eqnarray*}
where a residue of $1/2$ has to be taken into account at the origin in the case when $a=0$,
and ${\rm erf}\,\eta$ is the error function,
\begin{equation*}
  {\rm erf}\, \eta = \frac{2}{\sqrt\pi}\int_0^\eta e^{-t^2}\,\mathrm dt.
\end{equation*}
In particular,
\beq\lb{a29}
  L(x,t; 0) \approx \fr{1}{2} {\rm erfc} \left( \fr{x-t}{\sqrt{2t}} \right), \quad
  L(x,t; 1) \approx \fr{1}{\sqrt{2\pi t}} \exp \left[ - \fr{(x-t)^2}{2t} \right].
\eeq
Here ${\rm erfc} \, \eta = 1 -{\rm erf} \, \eta$ is the complementary error function.

\section*{References}
\begin{harvard}

\item Batchelor G K 1967 {\it An Introduction to Fluid Dynamics} (Cambridge Univ.)

\item Burgers J M 1920 On the resistance of fluids and vortex motion {\it Proceedings of the Koninklijke Akademie Van Wetenschappen Te Amsterdam} {\bf 23} 774-782

\item Chang C C, Su J Y and Lei S Y 1998 On aerodynamic forces for viscous compressible flow {\it Theort. Comput. Fluid Dyn.} {\bf 10} 71-90

\item Cole J D and Cook L P 1986
\textit{Transonic Aerodynamics}
(North-Holland)

\item Filon L N G 1926
The forces on a cylinder in a stream of viscous fluid
\textit{Proc. R. Soc. Lond. A}
\textbf{113} 7--27

\item Huang G C 1994 {\it Unsteady Vortical Aerodynamics: Theory and Applications} (Shanghai Jiaotong Univ., in Chinese)

\item Iima M 2008
A paradox of hovering insects in two-dimensional space
\textit{J. Fluid Mech.} \textbf{617} 207--229

\item Joukowski N E 1906
On annexed vortices
\textit{Proc. of Physical Section of the Natural Science Society}
\textbf{13} (2) 12--25

\item Lagerstrom P A 1964 {\it Laminar Flow Theory} (Princeton Univ.)

\item Lagerstrom P A, Cole J D and Trilling L 1949 Problems in the theory of viscous compressible fluids {\it Tech. Rep. 6. GALCIT Rept.}

\item  Landau L D and Lifshitz E M 1959
\textit{Fluid Mechanics}
(Pergamon Press)

\item  Lighthill M J 1952
On sound generated aerodynamically. I. General theory
\textit{Proc. R. Soc. Lond. A} \textbf{211} 564--587

\item Lighthill M J 1956 Viscosity Effects in Sound Waves of Finite
  Amplitude. (in {\it Surveys in Mechanics}, ed. Batchelor G K and Davies R M, Cambridge Univ.)

\item  Lighthill M J 1978
\textit{Waves in Fluids}
(Cambridge Univ.)

\item  Lighthill M J 1986
\textit{An Informal Introduction to Theoretical Fluid Mechanics}
(Clarendon Press)

\item Liu L Q 2016
\textit{Unified Theoretical Foundations of Lift and Drag in Viscous and Compressible External Flows}
Doctor thesis, Peking University (in Chinese)

\item  Liu L Q, Shi Y P, Zhu J Y, Su W D, Zou S F and Wu J Z 2014
Longitudinal-transverse aerodynamic force in viscous compressible complex flow
\textit{J. Fluid Mech.} \textbf{756} 226--251

\item  Liu L Q, Wu J Z, Shi Y P and Zhu J Y 2014
A dynamic counterpart of Lamb vector in viscous compressible aerodynamics
\textit{Fluid Dyn. Res.} \textbf{46} 061417

\item  Liu L Q, Zhu J Y and Wu J Z 2015
Lift and drag in two-dimensional steady viscous and compressible flow
\textit{J. Fluid Mech.} \textbf{784} 304--341

\item Mao F, Shi Y P and Wu J Z 2010 On a general theory for compressing process and aeroacoustics: linear analysis {\it Acta Mech. Sin.} {\bf 26} 355-364

\item Meng X G and Sun M 2015
Aerodynamics and vortical structures in hovering fruitflies
\textit{Phys. Fluids}
\textbf{27} 031901

\item  Noca F, Shiels D and Jeon D 1999
Lift and drag in two-dimensional steady viscous and compressible flow
\textit{J. Fluids Structures} \textbf{13} 551--578

\item Prandtl L 1918 Nachrichten von der gesellschaft der wissenschaften zu g\"{o}ttingen, mathematisch-physikalische klasse {\it Tragfl\"ugeltheorie. I. Mitteilung} 451-477

\item Saffman P G 1992 {\it Vortex Dynamics} (Cambridge Univ.)

\item von K\'arm\'an Th and Burgers J M 1935
General Aerodynamic Theory --- Perfect Fluids.
(In \textit{Aerodynamic Theory} Vol. II, ed. Durand W F, Dover)

\item  von K\'arm\'an Th and Sears W R 1938
Airfoil theory for non-uniform motion
\textit{J. Aero. Sci.} \textbf{5} (10) 379--390

\item Webster 1987
\textit{Webster's Ninth New Collegiate Dictionary}
(Merriam-Webster Inc. Publishers)

\item Wu J C 1981 Theory for aerodynamic force and moment in viscous flows {\it AIAA J.} {\bf 19} 432-441

\item Wu J C 1982 Problems of general viscous flow {In \it Developments in Boundary Element Method} (ed. P. K. Benerjee)

\item Wu J C 2005 {\it Elements of Vorticity Aerodynamics} (Tsinghua Univ. \& Springer)

\item Wu J Z, Lu X Y and Zhuang L X 2007 Integral force acting on a body due to local flow structures {\it J. Fluid Mech.} {\bf 576} 265-286

\item Wu J Z, Ma H Y and Zhou M D 2006 {\it Vorticity and Vortex Dynamics} (Springer)

\item Wu J Z, Ma H Y and Zhou M D 2015
\textit{Vortical Flows}
(Springer)

\item  Wu J Z, Pan Z L and Lu X Y 2005
A comparison of methods for evaluating time-dependent fluid dynamic forces on bodies, using only velocity fields and their derivatives
\textit{Phys. Fluids} \textbf{17} 098102

\item Wu T Y 1956 Small perturbations in the unsteady flow of a compressible, viscous and heat-conducting fluid {\it J. Math. Phys.} {\bf 35} 13-27

\item Xu C Y, Chen L W and Lu X Y 2010 Large-eddy simulation of the compressible flow past a wavy cylinder {\it J. Fluid Mech.} {\bf 665} 238-273

\item  Zhu J Y, Liu T S, Liu L Q, Zou S F and Wu J Z 2015
Causal mechanisms in airfoil-circulation formation
\textit{Physics of Fluids} \textbf{27} (12) 123601

\end{harvard}

\end{document}